\documentclass[12pt]{article}
\usepackage{graphicx}
\usepackage{dcolumn}
\usepackage{bm}
\usepackage{amsmath, amssymb}
\usepackage{color}
\usepackage{latexsym}

\begin{document} 

\setlength{\baselineskip}{18pt}
\begin{titlepage}
\begin{flushright}
OU-HET 637/2009 \\
YITP-09-51 
\end{flushright}

\vspace*{1.2cm}
\begin{center}
{\Large\bf Dirichlet Higgs in extra-dimension, consistent with 
electroweak data}
\end{center}
\lineskip .75em
\vskip 1.5cm

\begin{center}
{\large 
Naoyuki Haba$^{a,}$\footnote[1]{E-mail:\tt haba@phys.sci.osaka-u.ac.jp}, 
Kin-ya Oda$^{a,}$\footnote[2]{E-mail:\tt odakin@phys.sci.osaka-u.ac.jp}, 
and 
Ryo Takahashi$^{b,c,}$\footnote[3]{E-mail:\tt Ryo.Takahashi@mpi-hd.mpg.de}}\\

\vspace{1cm}

$^a${\it Department of Physics, Osaka University,
Osaka 560-0043, Japan}\\
$^b${\it Yukawa Institute for Theoretical Physics, Kyoto University,
Kyoto 606-8502, Japan}\\
$^c${\it Max-Planck-Institut f$\ddot{u}$r Kernphysik, Postfach 10 
39 80, 69029 Heidelberg, Germany}\\

\vspace*{10mm}
{\bf Abstract}\\[5mm]
{\parbox{13cm}{\hspace{5mm}

We propose a simple five-dimensional extension of the Standard Model (SM) without any Higgs potential nor any extra fields. A Higgs doublet lives in the bulk of a flat line segment and its boundary condition is Dirichlet at the ends of the line, which causes the electroweak symmetry breaking without Higgs potential. The vacuum expectation value of the Higgs is induced from the Dirichlet boundary condition which is generally allowed in higher dimensional theories. The lightest physical Higgs has non-flat profile in the extra dimension even though the vacuum expectation value is flat. As a consequence, we predict a maximal top Yukawa deviation (no coupling between top and Higgs) for the Brane-Localized Fermion and a small deviation, a multiplication of $2\sqrt{2}/\pi\simeq0.9$ to the Yukawa coupling, for the Bulk Fermion. The latter is consistent with the electroweak precision data within 90\% CL for $430\,\text{GeV}\lesssim m_{KK}\lesssim 500\,\text{GeV}$.
}}
\end{center}
\end{titlepage}

\section{Introduction}
The Standard Model (SM), with appropriate extension to take into account the observed neutrino masses, has passed all the experimental tests up to now.
In the model, all the masses for fermions and gauge bosons are solely from the Higgs mechanism.
Currently, the Higgs sector is the only missing part of the model, which is waiting to be tested at the CERN Large Hadron Collider (LHC).

The five dimensional Universal Extra Dimension (UED) model puts all the SM 
fields in the bulk of a compactified extra dimension $S^1/Z_2$, or equivalently 
of a line segment~\cite{Appelquist:2000nn,Appelquist:2002wb}. The electroweak 
symmetry breaking is caused by a bulk potential for the Higgs field. The LHC 
experiment might prove existence of the extra dimension if first few peaks of the Kaluza-Klein (KK) modes are discovered.

Another interesting phenomenological consequence from 
extra-dimensional theories is a top Yukawa deviation, which is a deviation of 
the Yukawa coupling between top and physical Higgs fields from the naive SM 
expectation. Such a deviation generically occurs in a 4-dimensional model too if
 there are multi-Higgs fields. Recently, it has pointed out that the deviation 
can be induced from effects of the brane localized Higgs potentials in the 
context of extra-dimensional theory \cite{Haba:2009uu,HOT3} even when there is 
only one Higgs doublet.\footnote{See Ref.\cite{Hosotani:2008by}
 for the top Yukawa deviation in a warped gauge-Higgs unification model.}

In this Letter, we point out that an extra dimensional model can
  predict a maximal top Yukawa deviation and the physical Higgs field can be as 
heavy as TeV without contradicting the electroweak precision measurements. We  
define the model compactified on a flat line segment with the Dirichlet boundary
 conditions (BCs) for a bulk Higgs field at the branes. In this model, the 
vacuum expectation value of the Higgs is induced from the Dirichlet BC which is 
generally allowed in higher dimensional theories. It is also shown that the 
resultant mass spectrum and interactions of the Higgs field are quite similar to
 the SM when we concentrate on the lowest modes in the KK expansions.

\section{Setup and model}
We consider a simple five-dimensional (5D) SM, compactified on a flat line segment, without adding any extra fields.

Let the SM gauge bosons and the Higgs doublet exist in the 5D flat space-time. The bulk-scalar kinetic action is given by
\begin{eqnarray}
S=-\int d^4x\int_{-L/2}^{+L/2}dz\left|D_M\Phi\right|^2,
\end{eqnarray}
where we write 5D coordinates as $x^M=(x^\mu,z)$ with $\mu=0,\dots,3$ and the extra dimension is compactified on a line segment $-L/2\leq z\leq L/2$. The five dimensional gauge covariant derivative is given as $D_M=\partial_M+igT^aW_M^a+ig'YB_M$ with $T^a=\sigma^a/2$ and $Y=1/2$ on the Higgs doublet field. Our metric convention is $(-++++)$. We also impose the KK parity, the reflection symmetry $z\to-z$ on the boundary conditions so that they are equal to each other at both boundaries, as in the UED model. 

The variation of the action is given by
 \begin{eqnarray}
  \delta S=\int d^4x\int_{-L/2}^{+L/2}dz
            \left[\delta\Phi(\mathcal{P}\Phi_X)
                  +\delta(z-\frac{L}{2})\delta\Phi(-\partial_z\Phi)
                  +\delta(z+\frac{L}{2})\delta\Phi(+\partial_z\Phi)\right],
 \end{eqnarray}
where $\mathcal{P}\equiv\Box+\partial_z^2$. The vacuum expectation 
value (vev) of the scalar field, $\Phi^c$, is determined by the action 
principle, $\delta S=0$, that is $\mathcal{P}\Phi^c=0$. The vev profile is fixed
 by the BCs. We have normally four choices of combination of Dirichlet and 
Neumann BCs at $z=\pm L/2$, namely
 \begin{eqnarray}
  (D,D),~~~(D,N),~~~(N,D),~~~\mbox{and}~~~(N,N),
 \end{eqnarray}
where the $D$ and $N$ means the Dirichlet and Neumann BCs, 
respectively. Difference choice of BC corresponds to different choice of theory.
 The theory is fixed once one chooses one of four conditions.

In this letter, we propose to take Dirichlet BCs for the Higgs field.
 The most general form of the Dirichlet BC is 
$\left.\delta\Phi\right|_{z=\pm L/2}=0$ and $\left.\Phi\right|_{z=\pm L/2}=(v_1,v_2)$ 
where $v_1$ and $v_2$ are free complex constants. Without loss of generality, we
 can always take a basis by an $SU(2)_L\times U(1)_Y$ field redefinition so that
 the boundary condition becomes
\begin{align}
\delta\Phi(x,z)|_{z=\pm L/2} &= 0, &
\left.\Phi(x,z)\right|_{z=\pm L/2} &= \begin{pmatrix}0\\ v\end{pmatrix},
\label{DD}
\end{align}
where $v$ is a real constant of mass dimension $[3/2]$. The BC~\eqref{DD} fixes 
the vev to be the fixed value $(0,v)$, while requiring the quantum fluctuation 
to be vanishing at the boundaries. The general solution of the equation of 
motion (EOM) takes the form 
$\Phi^c(z)\sim A+Bz$. The constants $A$ and $B$ are fixed by the BC~\eqref{DD} 
and the resultant vev profile becomes flat in the extra dimension
\begin{equation}
\label{vev}
\Phi^c(z)=\begin{pmatrix}0\\ v\end{pmatrix}.
\end{equation}
It is remarkable that the vev of Higgs field and flat profile in the 
extra-dimensional direction can be realized by taking the most general Dirichlet
 BC at the branes without contradiction to the action principle. The gauge 
symmetry is violated by this extra-dimensional BC and the gauge boson masses can
 be obtained. The constant vev makes profiles of the SM $W$ and $Z$ bosons to 
be flat too \cite{Haba:2009uu}.\footnote{Note that the resultant $Z$ 
and $W$ masses could be correct ones due to the custodial symmetry even under 
the presence of bulk Higgs mass. For simplicity, the bulk potential is assumed 
to be zero throughout this letter.} How about the profiles of 
quantum fluctuation modes of the Higgs field and the would-be Nambu-Goldstone 
(NG) bosons?

The Higgs doublet field $\Phi$ is KK-expanded as
\begin{align}
\Phi(x,z)=\begin{pmatrix}
         \sum_{n=0}^\infty
             f_n^{\varphi}(z)\varphi^{+(n)}(x) \\
          v+\frac{1}{\sqrt{2}}
          \sum_{n=0}^\infty\left[f_n^H(z)H^{(n)}(x)+if_n^\chi(z)\chi^{(n)}(x)\right]
         \end{pmatrix}
\end{align}
around the vev~\eqref{vev}.
Focusing on $H$,
the KK equation is given by
$\partial_z^2f_n^H(z)=-\mu_{Hn}^2f_n^H(z)$, which
has a solution
$f_n^H(z)=\alpha_n\cos(\mu_{Hn}z)+\beta_n\sin(\mu_{Hn}z)$.
The Dirichlet BC $\delta\Phi=0$ reads, for the quantum fluctuation,
\begin{eqnarray}
f_n^H(z)|_{z=\pm L/2}=0,
\end{eqnarray}
so that
the 5D profile of the quantum mode becomes
\begin{eqnarray}
f_n^H(z)=\left\{
     \begin{array}{ll}
      \sqrt{\frac{2}{L}}\cos\left(\frac{(n+1)\pi}{L}z\right) &
\text{for even $n$,} \\
      \sqrt{\frac{2}{L}}\sin\left(\frac{(n+1)\pi}{L}z\right) &
\text{for odd $n$.}
     \end{array}
    \right.
\end{eqnarray}
This means
that a flat zero-mode profile in the Neumann BC case is deformed to
the cosine function of
$f_0^H(z)=\sqrt{2/L}\cos(\pi z/L)$ through
the Dirichlet BC.
The $n$-mode Higgs mass
is calculated as
\begin{eqnarray}
\label{88}
m_{H^{(n)}}^2
=-\int_{-L/2}^{+L/2}dzf_n^H(z)\partial_z^2f_n^H(z)=\left(\frac{(n+1)\pi}{L}\right)^2,
\end{eqnarray}
which shows
the lowest ($n=0$) mode has
a KK mass $m_{\text{KK}}\equiv\pi/L$.
Note that $n=-1$ mode is vanishing.
This feature of KK scale Higgs mass is the specific result induced 
from the Dirichlet BC of Eq.(\ref{DD}). The mass is determined by only the 
compactification scale of extra-dimension unlike the SM.\footnote{We 
do not have a theoretical constraint on the magnitude of the Higgs mass from the
 discussions of perturbative unitarity as in the SM since the mass depends on 
the compactification scale but not on the Higgs self-coupling.}
Profiles of
$\chi^{(n)}$ and
$\varphi^{(n)\pm}$
are the same
as $H^{(n)}$.
Note that
this profile can be also realized by
introducing an extra {\it fake Higgs}
field $\phi$ at the boundaries, with the interaction with Higgs doublet as
$|\phi|^2 |\Phi|^2$, and taking a limit of
$|\langle \phi \rangle|\rightarrow \infty$.
This is the similar construction 
to the higgsless models.

The tree-level Higgs couplings,
$HHH$, $H\chi\chi$,
$H\varphi^+\varphi^-$,
$HHHH$, $HH\chi\chi$,
$HH\varphi^+\varphi^-$, $\chi\chi\chi\chi$,
$\chi\chi\varphi^+\varphi^-$,
and
$\varphi^+\varphi^-\varphi^+\varphi^-$
vanish,
since there is no Higgs potential.
It means that
longitudinal components of gauge bosons
$W_L$ and $Z_L$, only have the
gauge interactions.

We take Neumann BC,
\begin{equation}
\label{NN}
\partial_z A_\mu (z)|_{z=\pm L/2}=0 \;\;\;{\rm and}\;\;\;
A_z(z)|_{z=\pm L/2}=0,
\end{equation}
for the bulk gauge bosons.
Then the profile of the zero-mode gauge bosons
are flat and its KK masses are equal to the SM values
$m_Z^2=(g^2+g'{}^2)v^2/2$ and $m_W^2=g^2v^2/2$, respectively. The mass of the 
$n$th KK mode is given by
\begin{equation}
\label{wzmass}
m^2_{Z^{(n)},W^{(n)}}=m_{Z,W}^2+{n^2 \pi^2\over L^2}.
\end{equation}
Equations (\ref{88}) and (\ref{wzmass}) show
that bulk fields have the same magnitude of KK mass
(at the tree level).
We find that the mass of $n$-mode Higgs is same as the KK mass of gauge bosons 
with the KK number $n+1$ and the frequency of profile for the $n$-mode Higgs is 
also same as that of $n+1$-mode KK gauge bosons which are just results of the 
Dirichlet BC of the Higgs doublet.



Now we focus on the Higgs mechanism of the zero-mode gauge bosons. How is it 
possible to occur although the five dimensional fields $\varphi^\pm, \chi$ have
 no flat KK-mode while the lowest mode of the $W^\pm,Z$ are flat? The Higgs vev
 itself has a flat profile, and a linear combination of infinite KK modes of 
$\varphi^\pm$ and $\chi$ must have the flat profile, to be absorbed into 
$W^{(0)\pm}$ and $Z^{(0)}$, respectively, as the would-be NG bosons. This means
 that the longitudinal component, $W^{(0)\pm}_L$ ($Z^{(0)}_L$), is composed by 
a linear combination of $\varphi^{(n)\pm}$ ($\chi^{(n)}$). We speculate that, 
for example, $Z_L^{(0)}$ absorbs the following field having flat profile along 
the fifth direction except at the boundary,
\begin{eqnarray}
\chi \left(x,\pm {L\over2}\right)=0\;\; \hbox{and}\;\;
\chi (x,z)=\chi_{\text{NG}}(x) \;\;\left(-{L\over2}<z<{L\over2}\right),
\label{flat_like}
\end{eqnarray}
which can be realized by the superposition of
the infinite numbers of
$n$-modes of
$\chi^{(n)}$,
whose orthogonal linear combination is
the physical neutral pseudo-scalar. 
It is given by\footnote{See Ref. \cite{Haba:2010xz} for a detailed derivation.}
 \begin{eqnarray}
  \chi(x,z)=\chi_{\text{NG}}\sum_{m=0}^{\infty}\frac{4(-1)^m}{(2m+1)\pi}\cos\left(\frac{(2m+1)\pi}{L}z\right),
 \end{eqnarray}
where $n=2m$. In the same way,
a linear combination of $\varphi^{(n)\pm}$ is absorbed
into $W^{(0)\pm}$,
and its orthogonal linear combination becomes
physical charged scalar particle.
Thus,
the infinite numbers of
$n$-mode are required for the suitable
Higgs mechanism.

One possible question is:
``How should we treat
the 5D cutoff energy scale?''
There exist heavier KK modes than the cutoff scale,
and the completely flat profile of the would-be NG boson
cannot be obtained without such heavier KK modes.
However, a model with a cutoff $\Lambda$ is expected to have
an ambiguity of length scale of
${\mathcal O}(\Lambda^{-1})$ in general.
We would need an experimental
resolution finer than ${\mathcal O}(\Lambda^{-1})$
to distinguish this ambiguity
(for example, as the deviation from flat profile in above case),
and the ambiguity is negligible in the
low energy effective theory.

Let us comment on the KK parity.
It is known that the universal extra-dimensional (UED) model has
a KK parity conservation.
In our setup, the Dirichlet BC is imposed for the Higgs field
to take the same value on both $z=\pm L/2$ branes,
so that there arises a reflection symmetry.
This guarantees the conservation of the KK parity in
the gauge and Higgs sector.
The existence of the KK parity in the Lagrangian
depends on a fermion sector.
When the fermions are localized on the 4D branes
(brane-localized fermion (BLF) scenario),
the KK parity is broken in general.
On the other hand,
the KK parity is conserved
in a bulk fermion (BF) setup.
When KK parity exists,
the lightest KK particle with odd parity is stable, which
can be a candidate for a dark matter.

\section{Top Yukawa Deviation}

Now let us estimate
the top Yukawa deviation. 
This is a result from the non-flat profile of the physical Higgs field
in
the extra-dimension. 
We estimate the BLF scenario at first.
The Yukawa interaction for the top quark and the
Higgs boson is written as
\begin{eqnarray}
-\mathcal{L}_t&=&
\int_{-L/2}^{+L/2}dz\delta(z-\frac{L}{2}) 
y_{t,5}
               \left[v+f_0^H(z)\frac{H(x)}{\sqrt{2}}\right]\bar{t}(x)t(x)
\nonumber \\
            &=&y_{t,5}\left[v+f_0(L/2)\frac{H(x)}{\sqrt{2}}\right]
               \bar{t}(x)t(x).
\end{eqnarray}
The top quark mass $m_t$ and effective top coupling
in 4D $y_t$ can be
obtained as $m_t=y_{t,5}v$ and
$y_t=\frac{m_t}{v\sqrt{L}}=\frac{y_{t,5}}{\sqrt{L}},$ where we take
$v=v_{\mbox{\scriptsize EW}}/\sqrt{L}$.
On the other hand, the coupling
between the top quark and Higgs boson in 4D is given by
\begin{eqnarray}
\label{18}
y_{\bar{t}tH}=y_tf_0^H\left(\frac{L}{2}\right)
=y_t\sqrt{\frac{2}{L}}\cos\left(
\frac{\pi}{2}\right)=0.
\end{eqnarray}
This is the maximal top Yukawa deviation,
which can be hardly realized in other setups
({\it see for example,} Ref.~\cite{Hosotani:2008by}).
This maximal top Yukawa deviation
is the result of non-flat Higgs profile 
due to
the Dirichlet BC.

Next, in the case of the BF,
the Yukawa coupling between the top quark and the Higgs boson
is written by
\begin{eqnarray}
-\mathcal{L}_t&=&y_{t,5}\int_{-L/2}^{+L/2}dz
\left[v+f_0^H(z)\frac{H(x)}{\sqrt{2}}
                                     \right]\bar{t}(x,z)t(x,z).
\end{eqnarray}
Then
the ratio of the top Yukawa coupling in our model
to that of the SM, $r_{H\bar{t}t}$, is given by
\begin{eqnarray}
\label{rhtt}
r_{H\bar{t}t} &=& \frac{1}{\sqrt{L}}\int_{-L/2}^{L/2}dzf_0^H(z)
           =  \frac{2\sqrt{2}}{\pi}\simeq 0.90.
\end{eqnarray}
Therefore, the top deviation in the BF setup is 10\% decrease from
the SM.

\section{Higgs production and decay}

Let us consider the Higgs production and decay at LHC experiment. First, we 
show the Higgs production processes. (We can analyze higher KK Higgs production
in the same way.) The SM predicts that the gluon fusion with the top quark 
1-loop diagram ($WW$ fusion) dominates when 
$m_H\equiv m_{H^{(0)}}(=m_{KK})\leq 1$ TeV ($m_H\geq 1$ TeV). Since the Higgs 
has the same mass scale as the KK gauge bosons (and also KK fermions in the BF)
 in our setup, the Higgs mass must be large enough to be consistent with 
experiments. As shown in the next section, the KK scale must be larger than a 
few TeV (600 GeV \cite{Haisch:2007vb}) in BLF (BF) scenario. Anyhow, since the 
Yukawa couplings of Higgs with the top quark are modified, the processes for 
the Higgs production must be reanalyzed. 

In the BLF scenario, the gluon fusion process is strongly suppressed, 
since the Higgs is not coupled with the
top quark at the tree level. On the other hand, the $WW$ fusion process still exists, but the magnitude decreases because the coupling between $W$ and Higgs is modified as
\begin{eqnarray}
-\mathcal{L}_{WWH}&=&\frac{em_W}{2\sin\theta_W}
                   \frac{1}{2L}\int_{-L/2}^{+L/2}dz
                   f_0^H(z)f_0^{W^+}f_0^{W^-}
                   H(x)W^+(x)W^-(x)+\text{h.c.},
\end{eqnarray}
where $\theta_W$ is the Weinberg angle.
The ratio of the $WWH$
coupling in our 5D model
to the SM, $r_{WWH}$,
is estimated as
\begin{eqnarray}
r_{WWH}\equiv\frac{1}{\sqrt{L}}\int_{-L/2}^{+L/2}dzf_0^H(z).
\end{eqnarray}
Note that this ratio is the same as
$r_{H\bar{t}t}$ in Eq.~(\ref{rhtt}).
To conclude, the
Higgs production mainly occurs through
the $WW$ fusion in the BLF
scenario, which is decreased about 20\%
compared to the SM
due to the suppression by the factor
$r_{WWH}^2(=r_{H\bar{t}t}^2)\simeq0.81$.

Next let us consider the BF scenario. 
Around $m_H =600\sim 800$ GeV in the SM ($\sqrt{s}=14$ TeV), 
 the gluon fusion cross section is about 
 $10$ times larger than
 the $WW$ fusion. 
So how is in BF scenario
 where the coupling between the top quarks and Higgs
 boson is decreased by 10\% from the top Yukawa coupling?
It means that
the gluon fusion
process still exists,
but the
cross
section
(magnitude)
decreases 80\% (90\%) due to
$r_{H\bar{t}t}^2\simeq0.81$
($r_{H\bar{t}t}\simeq 0.9$)
compared
to
the
SM.
The $WW$ fusion process is the same as
the BLF scenario. Other production processes such as $q\bar{q}\rightarrow HW$,
$q\bar{q}\rightarrow HZ$, and $gg,q\bar{q}\rightarrow Ht\bar{t}$ are also suppressed. Therefore, comparing to the SM, the cross section for the Higgs production in the BF scenario decreases $81\%$ overall, while the branching ratios are not changed.

Finally let us show the Higgs decay. In the SM,
the process $H\rightarrow W^+W^-$
dominates
when $m_H>130$ GeV. In our setup, the decay width is estimated quite similarly 
as in the SM:
\begin{equation}
\Gamma_{H\rightarrow W^+W^-}\simeq {g^2 \over 64\pi}{m_H^3\over m_W^2} r_{WWH}^2.
\label{H_decay}
\end{equation}
In the next section, we will see that the Higgs mass must be larger than 6.8 
TeV (BLF scenario) and $430\,\text{GeV}\lesssim m_H\lesssim 500\,\text{GeV}$ 
(BF scenario).

Notice that in our setup $m_H=m_{\text{KK}}\gg m_W$, and the Higgs decay 
process would become equivalent to the process 
$H\to\varphi_{\text{NG}}^+\varphi_{\text{NG}}^-$ 
(where $\varphi_{\text{NG}}^\pm$ is the NG mode absorbed by the lowest mode of 
$W^\pm$), if the NG boson equivalence theorem is applicable in the mass 
spectrum. Since there is no Higgs potential in our model at all, $H$ cannot 
couple to $\varphi^+\varphi^-$, which means that $H$ would decay into $W^+W^-$ 
only through the transverse mode of $W^\pm$, that would lead to a suppressed 
decay width $\Gamma_{H\rightarrow WW}\simeq{g^2 \over 64\pi}m_H r_{WWH}^2$. But
 is it true? These would-be NG bosons $\varphi_{\text{NG}}^\pm$ are absorbed in
to $W^\pm$, and their wave function profiles are given by Eq.\eqref{flat_like}.
 It is the linear combination of all the higher KK modes, which means a lot of 
heavier KK modes (than the Higgs mass) are included. Obviously, the higher KK 
mode components ($n\geq2$) in $\varphi_{\text{NG}}$ (with the profile of 
Eq.\eqref{flat_like} and being absorbed into $W^\pm$) are heavier than the 
Higgs mass, thus, the physical decay process has been estimated as 
Eq.\eqref{H_decay}.

\section{Electroweak Precision Measurements}

Finally, let us estimate
constraints from electroweak (EW) precision measurements on this
setup.
As for the BLF scenario, 
 the present experimental data requires that the KK
 scale, that is equal to Higgs mass in our setup,
 must be larger than a 6.8 TeV 
 at the 95\% confidence level \cite{kk1}.
To fit
 $S$ and $T$ parameters \cite{stu2} in such a super-heavy Higgs
 scenario, 
 some extensions of the model, such as matter content,
 might be required.

\begin{figure}[t]
\begin{center}
\includegraphics[scale = 1.1]{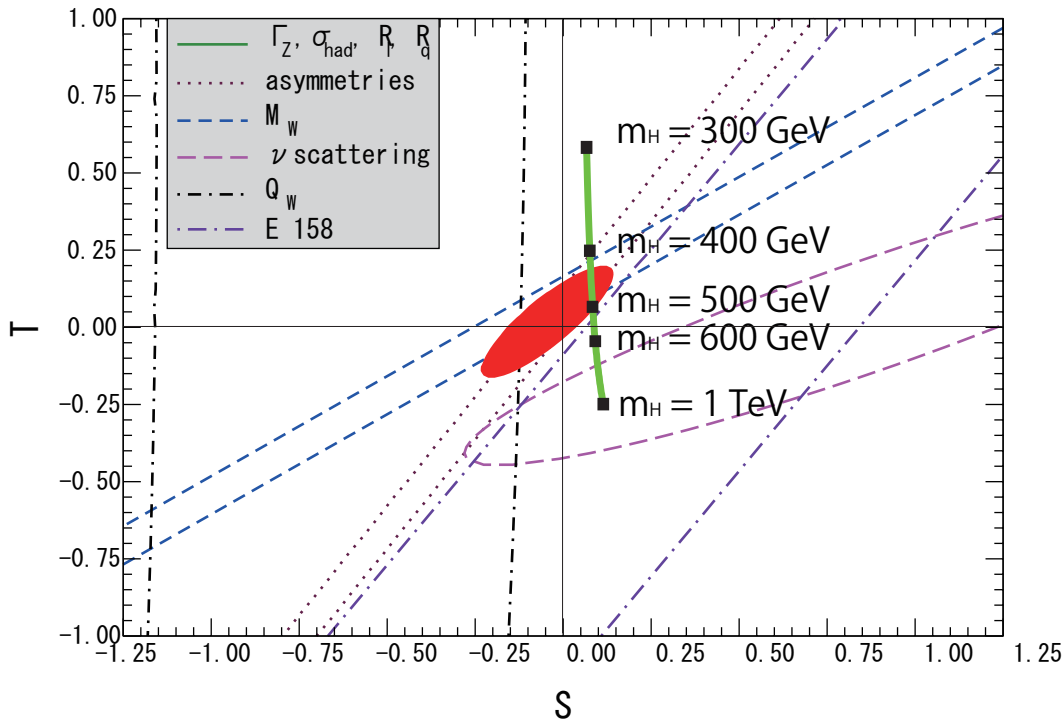}
\end{center}
\caption{$S$ and $T$ plot in this setup: 
Contours show $1\sigma$ constraints (39.35 \%) from various inputs except for 
the central one representing 90\% CL allowed by all data~\cite{pdg}.}
\label{fig1}
\end{figure}

In the BF case, we estimate $S$ and $T$ parameters defined as
$\alpha S\equiv 4e^2[\Pi_{33}^{\mbox{{\scriptsize new}}}{}'(0)
              -\Pi_{3Q}^{\mbox{{\scriptsize new}}}{}'(0)]$ and
$\alpha T\equiv\frac{e^2}{s^2c^2m_Z^2}[\Pi_{11}^{\mbox{{\scriptsize new}}}(0)
              -\Pi_{33}^{\mbox{{\scriptsize new}}}(0)]$,
where $s=\sin\theta_W$ and $c=\cos\theta_W$. $\Pi_{XY}(q^2)$ is the vacuum 
polarization and $\Pi_{XY}'(q^2)$ means $d\Pi_{XY}/dq^2$ at $q^2=0$. The 
$\Pi_{11}$ and $\Pi_{33}$ are represented by $\Pi_{11}=\frac{s^3}{e^2}\Pi_{WW}$
 and $\Pi_{33}=\frac{s^3}{e^2}[c^2\Pi_{ZZ}+2sc\Pi_{ZA}+s^2\Pi_{AA}]$, 
respectively. In our setup, $S$ and $T$ parameters are approximately estimated 
as \cite{Appelquist:2002wb,Gogoladze:2006br}
 \begin{eqnarray}
  S &\simeq& \frac{1}{6\pi}\log\left(\frac{m_H}{m_{H,\text{ref}}}\right)+\sum_{n=1}^\infty\frac{1}{4\pi}f_S^{\text{KK-top}}\left(\frac{m_t^2}{n^2m_{KK}^2}\right), \\
  T &\simeq& -\frac{3}{8\pi c^2}\log\left(\frac{m_H}{m_{H,\text{ref}}}\right)+\sum_{n=1}^\infty\frac{3m_t^2}{16\pi^2v_{\text{EW}}^2}\frac{1}{\alpha}f_T^{\text{KK-top}}\left(\frac{m_t^2}{n^2m_{KK}^2}\right),
 \end{eqnarray}
where $v_{\text{EW}}=174$ GeV, 
$m_H=m_{KK}$,\footnote{\label{absence_of_SM_Higgs}Here we approximate the 
absence of the SM Higgs by making its mass to be KK scale $m_H\to 
m_{H^{(0)}}=m_{KK}=\pi/L$, as the first KK Higgs $H^{(0)}$ has coupling to the 
SM zero-modes very close to the SM value, universally multiplied by 
$2\sqrt{2}/\pi\simeq 0.9$. Higher KK modes $H^{(n)}$ ($n\geq1$) are neglected 
as we will discuss in footnote~\ref{KK_Higgs_footnote}.} $m_{H,\text{ref}}$ is 
the reference Higgs mass taken as $m_{H,\text{ref}}=117$ GeV, and
 \begin{eqnarray}
  f_S^{\text{KK-top}}(z) &=& \frac{2z}{1+z}-\frac{4}{3}\log(1+z), \\
  f_T^{\text{KK-top}}(z) &=& 1-\frac{2}{z}+\frac{2}{z^2}\log(1+z).
 \end{eqnarray}
The first terms in both $S$ and $T$ parameters correspond to the absence of the
 SM Higgs contributions (as explained in footnote~\ref{absence_of_SM_Higgs}), 
and the second terms are the KK top ones. Since a contribution to $S$ and $T$ 
parameter from the KK Higgs modes are small at $m_{\text{KK}}\lesssim500$ 
GeV,\footnote{\label{KK_Higgs_footnote}In the UED model, contributions from the
 KK Higgs to $S$ parameter becomes dominant at $m_{\text{KK}}\gtrsim600$ GeV. 
However, such region of KK scale is excluded by the electroweak precision 
measurement at 90\% CL as shown in Fig.\ref{fig1}.} we drop the corresponding 
terms. We have not truncated the KK sum but performed it for infinite modes. 
Generically this is known to be a good strategy that does not spoil the five 
dimensional gauge symmetry at short distances. Notice that contributions from 
KK top loops, which are dominant contributions, are positive. The $(S,T)$ plot 
in this setup is presented in Fig.\ref{fig1}. We plot the parameters in a 
region of $300 \mbox{ GeV}\leq m_H\leq1\mbox{ TeV}$, and take the reference 
Higgs mass as $117$ GeV. We find that the first KK scale $m_H=\pi/L$ is 
constrained to be
\begin{align} 
430\,\text{GeV}\lesssim m_H\lesssim 500\,\text{GeV},
\end{align}
within 90\% CL. This numerical analysis is given with the KK sum until $11$th 
KK states, which is enough to evaluate the parameters, because contributions 
from higher KK modes to the parameters become negligibly tiny.

\section{Summary and discussions}

We have proposed the 5D SM with
the Higgs doublet and gauge bosons living in the bulk of the line segment.
We take a Dirichlet (Neumann) boundary condition for the Higgs (gauge) field. 
The vacuum expectation value of the Higgs is induced from the 
Dirichlet BC, which is generally allowed in higher dimensional theories, and 
the BC causes the electroweak symmetry breaking in this model. Under the simple
 BC, we have naturally obtained the non-flat profile of the lightest Higgs $H$. 
The mass of the physical Higgs boson is induced from the bulk 
quadratic terms and depends only on the compactification scale of the  
extra-dimension. We note that there is no Higgs self-coupling unlike the SM. In
 the BLF case, the maximal top deviation is realized, that is, top quark does 
not interact with physical Higgs boson. While in the BF case, 10\% deviation is
 predictive. We have shown that the Higgs decay width as large as its mass. The
 BF setup is consistent with $S$ and $T$ parameters. This model does not have 
unnatural large couplings nor any fine-tunings. 

Finally, we comment on unitarity in our model. Here, the gauge  
symmetry is violated by the extra-dimensional BC. However, the five  
dimensional gauge symmetry will be restored as an energy scale becomes  
much higher than the KK scale. We note that in several models of orbifold/boundary symmetry breaking, it has been shown that the longitudinal gauge boson scattering etc.\ are indeed unitarized by taking into account Kaluza-Klein mode contributions \cite{Hall,Chivukula0,Chivukula,Abe:2003vg,Chivukula1,Csaki,Abe}.
Therefore, it would be expected that  
the bulk gauge boson scattering is unitarized in such a region of our model (above  
the KK scale but lower than the five-dimensional cut-off scale) by  
taking account all the relevant KK modes. It would be worth studying  
this issue further.




\subsection*{acknowledgments}
We would like to thank T. Yamashita for very helpful discussions,
and also thank K. Hikasa and S. Matsumoto for useful discussions.
This work is partially supported by Scientific Grant by Ministry of
Education and Science, Nos.\ 20540272, 20039006, 20025004,
20244028, and 19740171. The work of RT is supported by
 the GCOE Program, The Next Generation of Physics, Spun from Universality and 
Emergence.

\end{document}